\documentclass[apj]{emulateapj}

\begin{document}

\shorttitle{X-rays from PSR J1023+0038}
\shortauthors{Bogdanov et al.}

\title{A \textit{Chandra} X-ray Observation of the Binary Millisecond Pulsar PSR J1023+0038}

\author{Slavko Bogdanov\altaffilmark{1,2}, Anne
  M. Archibald\altaffilmark{1}, Jason W.~T.~Hessels\altaffilmark{3,4},
  Victoria M. Kaspi\altaffilmark{1}, \\ Duncan Lorimer\altaffilmark{5},
  Maura A.~McLaughlin\altaffilmark{5}, Scott M.~Ransom\altaffilmark{6}, and Ingrid
  H.~Stairs\altaffilmark{7}}

\altaffiltext{1}{Department of Physics, McGill University, 3600 University Street, Montreal, QC H3A 2T8, Canada}

\altaffiltext{2}{Canadian Institute for Advanced Research Junior Fellow}

\altaffiltext{3}{Netherlands Institute for Radio Astronomy (ASTRON),
Postbus 2, 7990 AA Dwingeloo, The Netherlands}

\altaffiltext{4}{Astronomical Institute ``Anton Pannekoek'',
  University of Amsterdam, 1098 SJ Amsterdam, The Netherlands}

\altaffiltext{5}{Department of Physics, West Virginia University, 210E Hodges Hall, Morgantown, WV 26506, USA}

\altaffiltext{6}{National Radio Astronomy Observatory, 520 Edgemont Road,
Charlottesville, VA 22901, USA}

\altaffiltext{7}{University of British Columbia, 6224 Agricultural Road Vancouver, BC V6T 1Z1, Canada}

\begin{abstract}  
  We present a \textit{Chandra X-ray Observatory} ACIS-S variability,
  spectroscopy, and imaging study of the peculiar binary containing
  the millisecond pulsar J1023+0038. The X-ray emission from the
  system exhibits highly significant (12.5$\sigma$) large-amplitude
  (factor of 2--3) orbital variability over the five consecutive
  orbits covered by the observation, with a pronounced decline in the
  flux at all energies at superior conjunction. This can be naturally
  explained by a partial geometric occultation by the secondary star
  of an X-ray--emitting intrabinary shock, produced by the interaction
  of outflows from the two stars. The depth and duration of the
  eclipse imply that the intrabinary shock is localized near or at the
  surface of the companion star and close to the inner Lagrangian
  point.  The energetics of the shock favor a magnetically dominated
  pulsar wind that is focused into the orbital plane, requiring close
  alignment of the pulsar spin and orbital angular momentum axes.  The
  X-ray spectrum consists of a dominant non-thermal component and at
  least one thermal component, likely originating from the heated
  pulsar polar caps, although a portion of this emission may be from
  an optically-thin ``corona''.  We find no evidence for extended
  emission due to a pulsar wind nebula or bow shock down to a limiting
  luminosity of $L_X\lesssim3.6\times10^{29}$ ergs s$^{-1}$ (0.3--8
  keV), $\lesssim$$7\times10^{-6}$ of the pulsar spin-down luminosity,
  for a distance of 1.3 kpc and an assumed power-law spectrum with
  photon index $\Gamma=1.5$.
\end{abstract}

\keywords{pulsars: general --- pulsars: individual (PSR J1023+0038) --- stars: neutron --- X-rays: stars}

\section{INTRODUCTION}

The Galactic source FIRST J102347.6+003841 was discovered by
\citet{Bond02} who initially classified it as a magnetic cataclysmic
variable. Subsequent studies \citep{Thor05,Homer06} concluded that
this object is more likely a neutron star low-mass X-ray binary.  Its
true nature was finally unveiled with the Green Bank Telescope
discovery of 1.69-ms radio pulsations (Archibald et al.~2009),
establishing it as a compact binary containing a rotation-powered
millisecond pulsar (MSP).  The pulsar, PSR J1023+0038, is in a
circular 4.8-hour binary orbit with an optically identified
non-degenerate low-mass ($\sim$$0.2$ M$_{\odot}$) companion star
\citep{Arch09}. Perhaps more importantly, it is the first radio MSP
binary to have exhibited past evidence for an accretion disk. In 2001,
double-peaked emission lines characteristic of accretion disks,
accompanied by short-timescale flickering and a blue spectrum, were
observed at optical wavelengths \citep{Bond02, Szkody03}. There is no
evidence that a disk is currently present in the system
\citep{Wang09}. A plausible interpretation of this behavior is a very
recent transition from an accretion- to a rotation-powered neutron
star. If this is indeed the case, PSR J1023+0038 lends strong
observational support for the long-suspected evolutionary connection
between accreting MSPs in low-mass X-ray binaries \citep[][and
  references therein]{Wij10} and rotation-powered ``recycled'' MSPs
\citep{Alp82}.

In its present ``quiescent'' state, PSR J1023+0038 exhibits radio
eclipses at superior conjunction, when the secondary star is between
the pulsar and observer, as well as short, random and irregular
eclipses and dispersion measure variations at all orbital phases
\citep{Arch09}. Based on the orbital parameters of the binary and an
assumed $M=1.4$ M$_{\odot}$ pulsar, for the implied orbital
inclination of $\approx$46$^{\circ}$, the line-of-sight between the
pulsar and the Earth does not intersect the Roche lobe of the
companion at any point in the orbit \citep{Arch10}. Therefore, the
eclipses must be caused by material driven off the surface of the
secondary by the impinging pulsar wind. Such eclipses and dispersion
measure variations seen in similar MSP binaries both in globular
clusters \citep[see, e.g.,][for the case of PSR J1740--5340 in NGC
  6397]{DAm01} and in the field of the Galaxy \citep[see][concerning
  PSR J2215+51]{Hes11} can be attributed to the presence of matter
flowing out from the irradiated companion, as is likely the case for
J1023+0038.

The J1023+0038 system has been previously observed in X-rays with
\textit{XMM-Newton} \citep{Homer06} for 15 ks, with all instruments in
full imaging mode, and revisited in a target of opportunity
observation for 34.5 ks, with the EPIC pn instrument configured in
fast timing mode \citep{Arch10}. The latter observation, in addition
to detecting variability correlated with orbital phase, has revealed
compelling evidence for X-ray pulsations at the pulsar's rotational
period. The X-ray spectrum is relatively hard with power-law spectral
photon index $\Gamma\approx1-1.3$ and luminosity $L_X\approx 9 \times
10^{31}$ ergs s$^{-1}$ (0.5--10 keV), plus a possible thermal
component.  The nature of the J1023+0038 system and the evidence for
variability in the X-ray lightcurve favor interaction of the
relativistic particle wind from the pulsar with matter from the close
stellar companion as the source of the non-thermal radiation.  The
detection of X-ray pulsations by \citet{Arch10} in the
\textit{XMM-Newton} EPIC pn data implies that a portion of the
observed X-rays originates from the pulsar itself. Pulsed X-ray
emission from typical MSPs ($\dot{E}\approx10^{33}$ ergs s$^{-1}$) is
thermal in nature \citep{Zavlin06,Bog06,Bog09}, while in the less
common energetic MSPs ($\dot{E}\approx10^{36}$ ergs s$^{-1}$) the
emission is characteristically non-thermal (see, e.g., Rutledge et
al.~2004). Given that PSR J1023+0038 is moderately energetic, with
$\dot{E}=5\times10^{34}$ ergs s$^{-1}$, the true nature of the
pulsations is ambiguous \citep{Arch10}.

Most of the rotational energy lost by a pulsar is carried away by
winds comprised of relativistic particles and magnetic fields
\citep{Ken84}.  The winds shocked in the ambient medium produce pulsar
wind nebulae (PWNe) observable from the radio through $\gamma$-rays.
To date, extended X-ray emission has only been observed from two MSPs:
the canonical ``black widow'' PSR B1957+20 \citep{Fru88,Stap03} and
the nearby isolated PSR J2124--3358 (Hui \& Becker 2006).  As PSR
J1023+0038 moves through the ambient interstellar medium, it may also
generate sufficient ram pressure to confine the pulsar wind, resulting
in the formation of a bow shock, accompanied by a synchrotron-emitting
``cometary'' tail trailing behind the pulsar.  Given the strong
possibility that PSR J1023+0038 is a recently activated radio MSP, it
is important to search for an associated wind nebula. The spatial
structure of any X-ray tail along the proper motion direction could,
in principle, be employed as a ``timeline'' of the pulsar wind
activity. This is possible because any accretion onto a radio pulsar
would presumably extinguish the pulsar wind, which in turn would
result in interruptions in the injection of the energetic wind into
the interstellar medium.

Herein, we report on a \textit{Chandra X-ray Observatory} Advanced CCD
Imaging Spectrometer (ACIS) imaging spectroscopic observation of PSR
J1023+0038. This investigation offers further insight into the
properties of this remarkable binary.  The present work is outlined as
follows. In \S 2, we summarize the observations, data reduction and
analysis procedure. In \S3, we focus on the X-ray variability of PSR
J1023+0038. In \S 4 we present the spectroscopic analysis, while in
\S5 we conduct an imaging analysis. In \S6 we describe a simple
geometric model and in \S7 the physics of the intrabinary shock
emission. We offer conclusions in \S8.

\section{OBSERVATION AND DATA REDUCTION}

The \textit{Chandra} dataset was acquired on 2010 March 24 (ObsID
11075) in a single, uninterrupted 83.1-ks exposure. The pulsar
was placed at the nominal aim point of the back-illuminated ACIS-S3
CCD, configured in VFAINT telemetry mode. To minimize the effect of
photon pileup \citep{Davis01}, the detector was used in a custom
subarray mode with 256 pixel rows, starting from CCD row 385.

The data re-processing, reduction, and analysis were performed using
CIAO\footnote{Chandra Interactive Analysis of Observations, available
  at \url{http://cxc.harvard.edu/ciao/}} 4.2 and the corresponding
calibration products (CALDB 4.3.1).  Starting from the level 1 data
products of the ACIS-S observation, we first removed pixel
randomization from the standard pipeline processing to aid in the
search for extended emission near the point-source emission from the
pulsar. To search for large-scale diffuse emission, we applied the
background cleaning algorithm applicable to the VFAINT telemetry
mode. However, this procedure tends to reject real source counts for
relatively bright sources. Consequently, for the image subtraction,
spectroscopic, and variability analyses discussed below, the data
without the background cleaning applied were used.

%
%
\begin{figure}[!t]
\begin{center}
\includegraphics[width=0.42\textwidth]{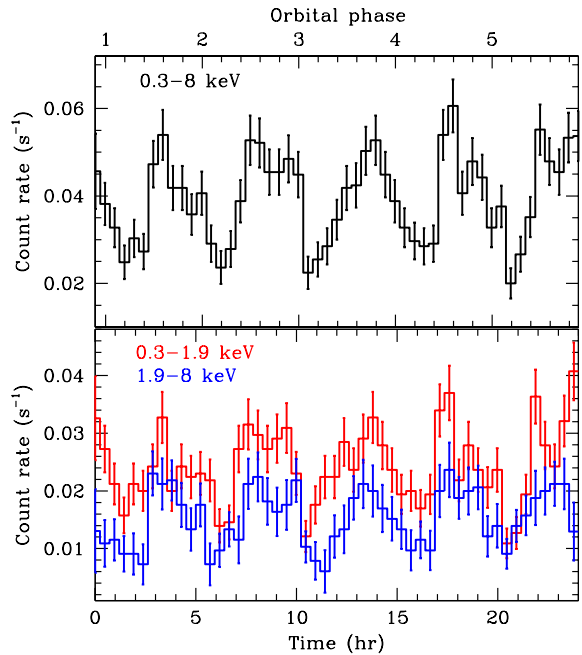}
\end{center}
\caption{\textit{Chandra} ACIS-S lightcurve of PSR J1023+0038 in the
  0.3--8 keV (\textit{Top panel}) and 0.3--1.9 keV and 1.9--8 keV
  bands (\textit{Bottom panel}). The five minima occur at
  $\phi_b\approx 0.1-0.4$ for each of the five full orbits covered by
  the observation. The orbital phase is defined based on the radio
  pulsar timing convention, in which superior conjunction (when the
  secondary star is between the pulsar and observer) occurs at
  $\phi_b=0.25$. The background contributes only 0.1\% to the total
  count rate.}
\end{figure}

For the spectroscopic and variability analyses, we extracted the
emission from a circular region of radius 2$\arcsec$ that encloses
$\gtrsim$90\% of the total source energy at 1.5 keV. The pulsar net
count rate in this region is $0.0394\pm0.0007$ counts s$^{-1}$ (0.3--8
keV). To permit spectral fitting in XSPEC\footnote{Available at
  \url{http://heasarc.nasa.gov/docs/xanadu/xspec/index.html}.}
12.6.0q, the extracted source counts in the 0.3--8 keV range were
grouped in energy bins so as to ensure at least 15 counts per bin. The
background was taken from three source-free regions in the image
around the pulsar.  For the variability analysis, the photon arrival
times were translated to the solar system barycenter using the CIAO
tool {\tt axbary} assuming the DE405 JPL solar system ephemeris and
the pulsar position derived from radio timing \citep{Arch09}.  We note
that the 3.2-s time resolution of the ACIS-S precludes the study of
pulsations at the pulsar spin period.

\section{X-ray Orbital Variability}

%
%
\begin{figure}[!t]
\begin{center}
\includegraphics[width=0.42\textwidth]{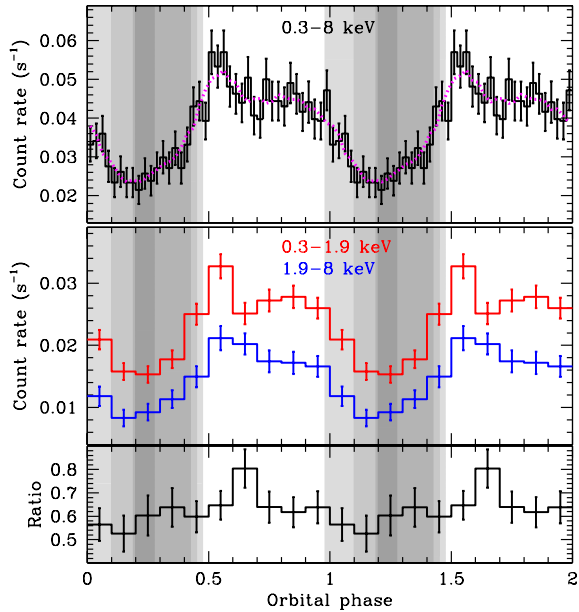}
\end{center}
\caption{(\textit{Top panel}) \textit{Chandra} ACIS-S emission in the
  0.3--8 keV range from PSR J1023+0038 folded at the 4.8-hour binary
  period. 
The dotted magenta line is the lightcurve smoothed using a
four-point moving average. 
(\textit{Middle panel}) Lightcurves in
  the 0.3--1.9 keV (\textit{red}) and 1.9--8 keV (\textit{blue})
  bands. (\textit{Bottom panel}) Ratio of the lightcurves in the
  middle panel, with the 1.9--8 keV divided by the 0.3--1.9 keV count
  rate.  For reference, the grey bands represent the approximate
  portions of the orbit where the pulsar undergoes radio eclipses at
  0.35, 0.7, 1.6, and 3 GHz (from lightest to darkest grey,
  respectively). Two cycles are shown for clarity.}
\end{figure}

The study by \citet{Arch10} uncovered large amplitude X-ray
variability from J1023+0038. However, due to the short exposure,
it was not possible to establish whether the flux modulations were
truly periodic. We have used the radio timing ephemeris of PSR
J1023+0038 \citep{Arch09} to fold the barycentered \textit{Chandra}
source photons at the binary period. As the observation covers over
five binary orbits, we can definitively confirm that the X-ray flux is
modulated at the binary period (Figures 1 and 2). The variations are
characterized by a factor of $\sim$2-3 decline in the photon count
rate at superior conjunction ($\phi_b\approx0.25$).  A Kuiper test
\citep{Pal04} on the folded, unbinned lightcurve, weighted to account
for the non-uniform exposure across the orbit, indicates a
$6\times10^{-36}$ (12.5$\sigma$) probability that photons being drawn
from a constant distribution would exhibit this level of
non-uniformity. Statistically, the individual lightcurves from the
five binary orbits are consistent with having the same shape.  There
is no indication of statistically significant spectral variability
throughout the orbit (bottom panel of Figure 2).  As discussed in \S6,
the X-ray modulations can be plausibly interpreted as being due to a
partial geometric occultation of an intrabinary shock by the secondary
star.

\begin{deluxetable*}{lcccc}
\tabletypesize{\small} 
\tablecolumns{10} 
\tablewidth{0pc}
\tablecaption{Summary of Spectral fits for PSR J1023+0038.}

\tablehead{\colhead{} & \colhead{Total} &
  \colhead{$\phi_{b,1}$} & \colhead{$\phi_{b,2}$} & \colhead{Joint} \\
  \colhead{Model\tablenotemark{a}} & \colhead{ } &
  \colhead{$(0.0-0.5)$} & \colhead{$(0.5-1.0)$} &
  \colhead{$\phi_1+\phi_2$} } \startdata
\textbf{Power-law} & 	&	&	&	\\
\hline
$N_{\rm H}$ ($10^{20}$ cm$^{-2}$) 	  & $<0.15$	& $<0.23$	& $<0.48$	&  $<0.17$	\\
$\Gamma$	          & $1.19^{+0.03}_{-0.03}$	& $1.29^{+0.05}_{-0.05}$	& $1.14^{+0.04}_{-0.03}$	& $1.29^{+0.05}_{-0.05}/1.14^{+0.04}_{-0.04}$ \\
$F_X$ ($10^{-13}$ ergs cm$^{-2}$ s$^{-1}$)\tablenotemark{c} & $3.80^{+0.10}_{-0.09}$  	& $2.78^{+0.11}_{-0.11}$	& $4.66^{+0.15}_{-0.12}$	& $2.77^{+0.11}_{-0.11}/4.66^{+0.15}_{-0.15}$	\\
$\chi^2_{\nu}$/dof       & $0.99/163$ & $1.19/71$	& $1.20/102$	&  $1.19/174$	\\
\hline
\textbf{Power-law + H atmosphere (NSA)\tablenotemark{b}}	& 	&	&	&	\\
\hline
$N_{\rm H}$ ($10^{20}$ cm$^{-2}$)	  & $<5.4$  & $7.9^{+3.2}_{-5.5}$ & $<7.9$	& $7.3^{+2.4}_{-4.4}$	\\
$\Gamma$	  & $1.00^{+0.05}_{-0.08}$	& $1.09^{+0.07}_{-0.06}$	& $1.00^{+0.10}_{-0.10}$	& $1.09^{+0.07}_{-0.07}/1.08^{+0.08}_{-0.07}$ 	\\
$T_{\rm eff}$ ($10^6$ K)       & $0.76^{+0.18}_{-0.25}$	& $0.44^{+0.03}_{-0.07}$ & $1.05^{+0.50}_{-0.43}$ 	& $0.55^{+0.09}_{-0.06}$ 	\\
$R_{\rm eff}$ (km)      &  $2.5^{+8.2}_{-1.1}$        & $16.3^{+7.8}_{-10.6}$ 	& $1.1^{+4.4}_{-1.0}$ 	& $8.3^{+9.2}_{-5.1}$ 	\\
Thermal fraction\tablenotemark{d}      &	$0.06^{+0.04}_{-0.04}$ & $0.21^{+0.09}_{-0.08}$ 	& $0.06^{+0.06}_{-0.05}$ 	& $0.17^{+0.07}_{-0.07}/<0.06$ 	\\
$F_X$ ($10^{-13}$ ergs cm$^{-2}$ s$^{-1}$)\tablenotemark{c}	& $4.15^{+0.15}_{-0.12}$ 	& $3.71^{+0.30}_{-0.28}$ & $4.97^{+0.21}_{-0.21}$ & $3.49^{+0.21}_{-0.21}/4.86^{+0.13}_{-0.13}$	\\
$\chi^2_{\nu}/$dof       & $0.84/161$ & $0.85/69$ & $1.18/100$	& $1.06/172$ \\
\hline
\textbf{Power-law + MEKAL}\tablenotemark{e}	& 	&	&	&	\\
\hline
$N_{\rm H}$ ($10^{20}$ cm$^{-2}$)	  & $<0.54$  & $<0.72$ & $<1.3$	& $<0.67$	\\
$\Gamma$	  & $1.07^{+0.04}_{-0.03}$	& $1.11^{+0.07}_{-0.05}$	& $1.05^{+0.05}_{-0.05}$	& $1.13^{+0.06}_{-0.06}/1.03^{+0.04}_{-0.04}$ 	\\
$kT$ (keV)       & $0.22^{+0.18}_{-0.18}$	& $0.19^{+0.02}_{-0.03}$ & $0.26^{+0.05}_{-0.05}$ 	& $0.22^{+0.02}_{-0.02}$ 	\\
Thermal fraction\tablenotemark{d}      &	$0.033^{+0.033}_{-0.032}$ & $0.06^{+0.05}_{-0.05}$ 	& $<0.07$ 	& $0.17^{+0.07}_{-0.07}/0.11^{+0.05}_{-0.05}$ 	\\
$F_X$ ($10^{-13}$ ergs cm$^{-2}$ s$^{-1}$)\tablenotemark{c}	& $4.09^{+0.09}_{-0.09}$ 	& $3.03^{+0.10}_{-0.11}$ & $4.81^{+0.15}_{-0.15}$ & $2.97^{+0.09}_{-0.09}/4.86^{+0.13}_{-0.13}$	\\
$\chi^2_{\nu}/$dof       & $0.84/161$ & $0.86/69$ & $1.16/100$	& $1.06/172$
\enddata 

\tablenotetext{a}{All uncertainties and upper limits quoted are 1$\sigma$.}

\tablenotetext{b}{For the NSA model, a $M=1.4$ M$_{\odot}$, $R=10$ km
  neutron star is assumed. The effective emission radius as measured
  at the stellar surface, $R_{\rm eff}$, was calculated assuming a distance of 1.3 kpc \citep{Arch09}.}

\tablenotetext{c}{Unabsorbed X-ray flux (0.3--8 keV) in units of
  $10^{-13}$ ergs cm$^{-2}$ s$^{-1}$.}

\tablenotetext{d}{Fraction of unabsorbed flux from the thermal component in the 0.3--8 keV band.} 

\tablenotetext{e}{For the MEKAL model, solar abundances are assumed.}

\end{deluxetable*}

\section{Spectroscopy}

\subsection{Total Spectrum}

As found by \citet{Homer06} and \citet{Arch10} based on the set of
\textit{XMM-Newton} observations, the phase-integrated
\textit{Chandra} X-ray spectrum of PSR J1023+0038 is well described by
a pure absorbed power-law, while it is poorly fitted by a single
thermal (blackbody or neutron star hydrogen atmosphere)
model. Moreover, \citet{Arch10} found that a slight improvement in the
fit quality was obtained with the addition of a thermal (neutron star
atmosphere) component. Based on this, in the spectral analysis of the
\textit{Chandra} data we consider both a one-component absorbed
power-law model and a composite absorbed power-law plus NS atmosphere.
We use the hydrogen atmosphere NSA model \citep{Zavlin96} over a
blackbody because an atmospheric layer is expected at the surface of a
MSP given the standard formation scenario of MSPs, involving accretion
of a substantial amount of matter. In addition, the thermal pulsations
from the nearest known MSPs require the presence of a light-element
atmosphere \citep{Bog07,Bog09}.  In principle, a portion of the
observed X-ray emission from the J1023+0038 system could also
originate from a thermal plasma within or around the binary, possibly
from the active corona of the secondary star or the plasma responsible
for the radio eclipses. To explore this possibility, we also conduct
fits with a model consisting of a power-law plus MEKAL thermal plasma
model in XSPEC. The MEKAL model considers an emission spectrum from
hot diffuse gas including line emissions from several elements given a
set of metal abundances \citep{Mewe85,Mewe86,Lie95}. The best-fit
parameters from the three different models are summarized in column 2
of Table 1.

Both the pure power-law and power-law plus thermal component yield
statistically acceptable fits. However, an F-test indicates that the
probability the composite model is a better fit purely by chance is
only $1.2\times10^{-6}$, suggesting that the addition of the thermal
component is warranted by the data. By analogy with most MSPs detected
in X-rays \citep{Zavlin06,Bog06,Bog09}, PSR J1023+0038 should have
heated magnetic polar caps and hence a thermal component in the
predominantly non-thermal spectrum.  The evidence for relatively broad
pulsations at the pulsar period found by \citet{Arch10} in the
\textit{XMM-Newton} EPIC pn data are consistent with a portion of the
observed X-rays originating from the neutron star polar caps.  The
implied thermal luminosity of $6\times10^{30}$ ergs s$^{-1}$ and
$\dot{E}$ to $L_X$ conversion efficiency ($\sim$$10^{-4}$) are
comparable to those of most nearby \citep{Zavlin06,Bog09} and globular
cluster MSPs \citep{Bog06,Bog11}.

As in \citet{Arch10}, in most of the fits, the hydrogen column density
along the line of sight is consistent with zero. This is not
surprising, given that the column density along the line of sight and
to the edge of the Galaxy is only $N_{\rm H}\approx1\times10^{20}$
cm$^{-2}$.  Since such a low $N_{\rm H}$ only affects the spectrum
below $\sim$0.3 keV, the fits of the \textit{Chandra} spectrum are
nearly insensitive to this parameter.

%
%
\begin{figure}[!t]
\begin{center}
\includegraphics[width=0.42\textwidth]{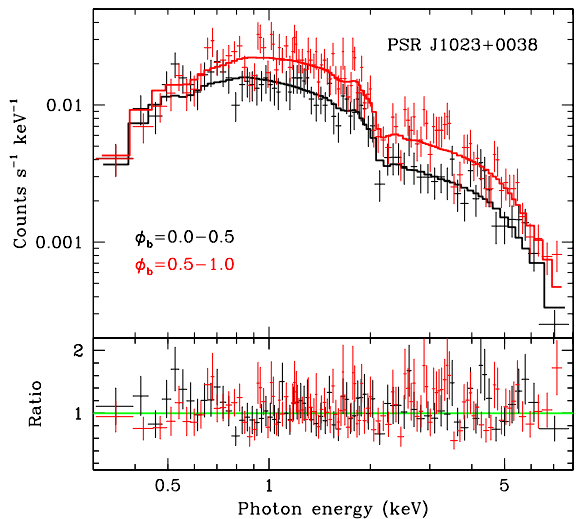}
\end{center}
\caption{\textit{Chandra} ACIS-S X-ray spectra of PSR J1023+0038 for
  orbital phases $\phi_b=0.0-0.5$ (\textit{black}) and
  $\phi_b=0.5-1.0$ (\textit{red}), fitted with a model consisting of
  an absorbed power law plus neutron star H atmosphere. The
  lower panel shows the ratio between the observed flux and the
  predicted model flux for each energy bin. See text and Table 1 for
  best-fit parameters.}
\end{figure}

\subsection{Orbital Phase-resolved Spectrum}

The ample number of photons (3280 within 2$\arcsec$ of the pulsar)
makes it possible to consider the continuum emission from J1023+0038
from two portions of the orbit: $\phi_b=0.0-0.5$ (at the X-ray
minimum) and $\phi_b=0.5-1.0$ (Figure 3). We fit the two spectra both
separately and jointly. In the joint fits, we tied $N_{\rm H}$ and
both the temperature and effective radius of the thermal component in
all cases since the radiation from the pulsar is not expected to
exhibit any dependence on orbital phase.  The results of the orbital
phase-resolved spectral fits are summarized in columns 3--5 of Table
1.

Although statistically a pure power-law spectrum provides a good
description of the total phase-averaged X-ray spectrum, it yields only
marginally acceptable fits of the individual spectra at orbital phases
$0-0.5$ and $0.5-1$, with $\chi_{\nu}^2\approx1.2$ in both cases. The
same is true for the joint fit of the two phase intervals.  In the
individual and joint fits, the inferred spectral photon indices are
consistent with being the same. This is in agreement with the finding
in \S3 regarding the absence of orbital phase-dependent spectral
variability although deeper observations are necessary to confirm
this.

In general, the spectral fits tend to favor the non-thermal plus
thermal model over the purely non-thermal model.  It should be noted,
however, that the derived temperatures and radii for the NSA model are
considerably different between the total, individual and joint
spectral fits.  One possible explanation is the presence of a
multi-temperature thermal spectrum, arising due to non-uniform heating
across the face of the magnetic polar caps. Such thermal emission is
seen in several nearby MSPs \citep{Zavlin06,Bog09}. The variation in
the dominant non-thermal component as a function of orbital phase
could result in the fits being more sensitive to the different
temperatures such that at the X-ray minimum the emission from the
cooler portions of the polar cap(s) becomes more prominent.
Alternatively, the differences in inferred temperature could be caused
by the presence of both polar cap and ``coronal'' thermal plasma
emission, especially if the coronal emission is also modulated at the
binary period.  Detailed spin and orbital phase-resolved spectroscopic
observations are required to unambiguously disentangle the various
spectral components.

\section{Imaging Analysis}

As noted in \S1, evidence of past pulsar wind activation and accretion
episodes in PSR J1023+0338 may, in principle, be manifested through
irregularities or discontinuities in any nebular X-ray emission
associated with the pulsar.  The expected stand-off angular distance
for a bow shock around a pulsar moving supersonically through the
ambient medium can be estimated using the expression
$\theta=n^{-1/2}\mu^{-1}\dot{E}^{1/2}D^{-2}$ \citep[][and references
  therein]{Kar08}, where $n$ is the mean density of the ambient gas,
$\mu$ is the pulsar proper motion, $\dot{E}$ the pulsar spin-down
luminosity, and $D$ the distance. For a distance of $D=1.3$ kpc
\citep{Arch09}, $\mu=18$ mas yr$^{-1}$, $\dot{E}=5\times10^{34}$ ergs
s$^{-1}$, and $n\approx0.1$ cm$^{-3}$ expected for warm neutral
ambient gas \citep[see, e.g.,][]{Chat02}, we obtain a stand-off
distance of $\sim$4$\arcsec$.

%
%
\begin{figure}[t]
\begin{center}
\includegraphics[width=0.42\textwidth]{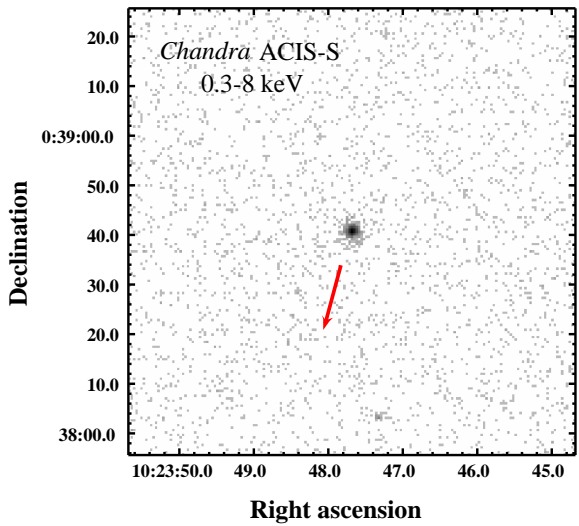}
\end{center}
\caption{\textit{Chandra} ACIS-S3 $1.5\arcmin\times1.5\arcmin$ image
  in J2000 coordinates centered on the X-ray counterpart to PSR
  J1023+0038 in the 0.3--8 keV band. The image is binned at the
  intrinsic detector resolution of $0.5\arcsec$. The
  grey scale corresponds to counts increasing logarithmically from
  white to black. The red arrow shows the direction of the pulsar
  proper motion (Deller et al.~in preparation). Pixel randomization
  has been removed from the pipeline processing.}
\end{figure}

%
%
\begin{figure}[t]
\begin{center}
\includegraphics[width=0.25\textwidth]{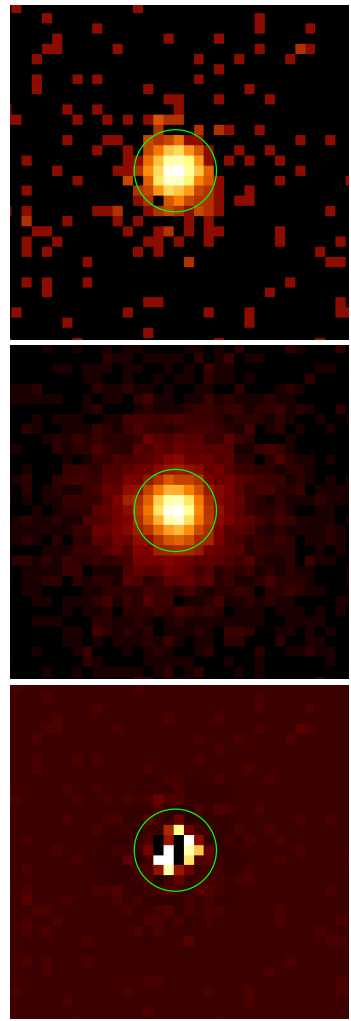}
\end{center}
\caption{(\textit{Top}) \textit{Chandra} ACIS-S 0.3--8 keV image of
  the X-ray counterpart to PSR J1023+0038 from ObsID 11075. The color
  scale corresponds to counts increasing logarithmically from black to
  white. The green circles in all three images are of radius
  2$\arcsec$ and are centered on the X-ray source position.
  (\textit{Middle}) Average of 100 simulated point spread functions of
  PSR J1023+0038 with ChaRT and MARX assuming the observing parameters
  of ObsID 11075 and the spectrum of the pulsar as inferred from the
  formal spectral fits. The color scale is the same as in the top
  panel. (\textit{Bottom}) The resulting difference image (observed
  minus simulated). The artifacts within 1$\arcsec$ are most likely
  caused by imperfections in the simulated PSF core. Beyond the PSF
  core, the residuals are consistent with being solely due to
  statistical fluctuations.}
\end{figure}

Figure 4 shows the image of the ACIS-S exposure centered on PSR
J1023+0038 in the $0.3-8$ keV band. The binary is the brightest source
in the image at the {\tt wavdetect}-reported position of
$\alpha=10^{\rm h}23^{\rm m}47\fs68$,
$\delta=+00^{\circ}38\arcmin40\farcs99$. This differs from the optical
position $\alpha=10^{\rm h}23^{\rm m}47\fs67$,
$\delta=+00^{\circ}38\arcmin41\farcs2$ given in the NOMAD catalog
\citep{Zach04} by just $+0.15\arcsec$ and $-0.21\arcsec$ in right
ascension and declination, respectively, well within the typical
uncertainty in the absolute astrometry of
\textit{Chandra}\footnote{See
  \url{http://cxc.harvard.edu/cal/ASPECT/celmon/}}.  It is evident that
beyond $\sim$2$\arcsec$ of the pulsar, there is no significant excess
of photons surrounding the pulsar, as might be expected from a pulsar
wind nebula, nor in the direction opposite the pulsar proper motion as
expected for a fast-moving pulsar.

PSR B1957+20 ($D=2.5$ kpc, $\dot{E}=10^{35}$ ergs s$^{-1}$) exhibits
an X-ray tail of length $\sim$16$\arcsec$ in the direction opposite
the pulsar proper motion \citep{Stap03}. Extracting the counts from a
rectangular region $20\arcsec \times 5\arcsec$ trailing J1023+0038
yields no appreciable excess of counts above the background count rate
near the pulsar position, which is $2\times10^{-6}$ counts s$^{-1}$
arcsec$^{-2}$ (0.3--8 keV). Assuming a representative power-law
spectrum from the current sample of X-ray PWNe \citep{Kar08} with
$\Gamma=1.5$, the conservative limit on the wind nebula luminosity is
$\sim$$3.6\times10^{29}$ ergs s$^{-1}$ for a distance of $1.3$ kpc.

%
%
\begin{figure}[t]
\begin{center}
\includegraphics[width=0.42\textwidth]{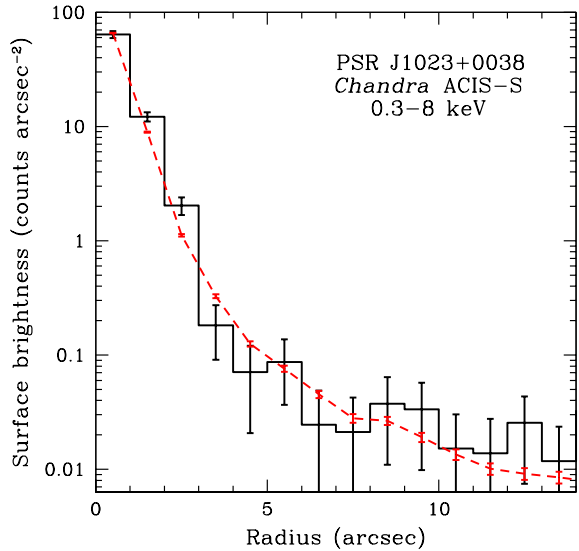}
\end{center}
\caption{Radial profile of the observed \textit{Chandra} ACIS-S point spread
  function of PSR J1023+0038 (\textit{histogram}) and the average of 100
  simulated PSFs with ChaRT and MARX (\textit{red dashed line}).}
\end{figure}

To investigate the presence of small-scale ($\lesssim2\arcsec$)
diffuse emission that could potentially arise due to a very compact
pulsar wind nebula, we simulated a set of 100 observations of PSR
J1023+0038 with the ChaRT\footnote{The Chandra Ray Tracer, available
  at \url{http://cxc.harvard.edu/soft/ChaRT/cgi-bin/www-saosac.cgi}}
and MARX 4.5\footnote{Available at
  \url{http://space.mit.edu/cxc/marx/index.html}.} software packages,
based on the parameters of the ACIS-S observation and the spectral
shape of the X-ray emission as determined from the spectral fits (see
\S 4). We subtracted the resulting average synthetic point spread
function from the observed image and examined the significance of any
residual emission relative to the background level (Figure 5).  Within
$\sim$1$\arcsec$ of the center, the difference image exhibits
significant azimuthally asymmetric residuals (both positive and
negative). One possible cause could be a slight offset between the
location of the source centroid reported by {\tt wavdetect} and the
true source position. To test this possibility, we have simulated
images with a range of displacements along the azimuthal direction in
which the residuals are most pronounced in an effort to reduce the
observed artifacts. However, in all instances there is no improvement
in the difference image. Therefore, it appears more likely that the
artifacts are due to imperfections in the modeling of the
High-Resolution Mirror Assembly optics \citep{Juda10}\footnote{See
  also
  \url{http://hea-www.harvard.edu/\~juda/memos/\\HEAD2010/HEAD2010\_poster.html}
  and
  \url{http://cxc.harvard.edu/cal/Hrc/PSF/acis\_psf\_2010oct.html}}.
Aside from the discrepancies in the central pixels of the PSF, the
spatial distribution of events beyond $\sim$1$\arcsec$ of the pulsar
position is consistent with that of a point source (Figure 6).

\section{A Geometric Model of the Intrabinary Shock}
The morphology and spectral properties of the folded X-ray lightcurve
in Figure 2 allow us to make important qualitative statements
regarding the properties of the source of the non-thermal X-ray
emission in the J1023+0038 system. For instance, the lack of
significant accompanying spectral variability indicates that the X-ray
variability is not due to photoelectric absorption. An enhanced
$N_{\rm H}$ during eclipse phases would diminish the softest portion
of the spectrum causing an apparent hardening of the emission.
Instead, a uniform decline at all energies is observed.  Thus, a more
natural interpretation is obstruction of our view of the
X-ray-emitting region by the companion star.  Although it is difficult
to pinpoint the exact start and end, the approximate duration of the
X-ray eclipse ($\sim$0.1--0.4 in orbital phase) implies that the
source of the X-ray emission is much closer to the secondary star
(presumably near or at the inner Lagrangian [L1] point) than to the
pulsar. For reference, for an assumed pulsar mass of 1.4 M$_{\odot}$,
which implies an orbital inclination of $46^{\circ}$, the L1 point is
occulted for 32\% of the orbit ($\phi_b=0.09-0.41$). The rounded shape
of the X-ray minimum and the gradual ingress and egress indicate that
the region is gradually occulted, implying it is extended (of order the
size of the star).
The factor of $\sim$2-3 decline in the X-ray flux implies that at
$\phi_b\sim0.25$ the secondary star blocks our line of sight to at
least half of the emission region.

To verify these statements, we have constructed a three-dimensional
geometric model of the binary assuming a 1.4 M$_{\odot}$ pulsar and
$i=46^{\circ}$ plus the orbital parameters measured from radio timing
\citep{Arch09}. Using this model, we generate lightcurves to compare
against the \textit{Chandra} data.  For the entire range of orbital
phases, we determined the fraction of the emission region under
consideration that was occulted by the secondary star as viewed by the
observer from a $46^{\circ}$ angle.  Based on this, we determined the
observed flux relative to when the region is in full view.  The
resulting lightcurves were subsequently fitted to the data, with a
flux scaling and DC offset as the only free parameters.

%
%
\begin{figure}[!t]
\begin{center}
\includegraphics[width=0.42\textwidth]{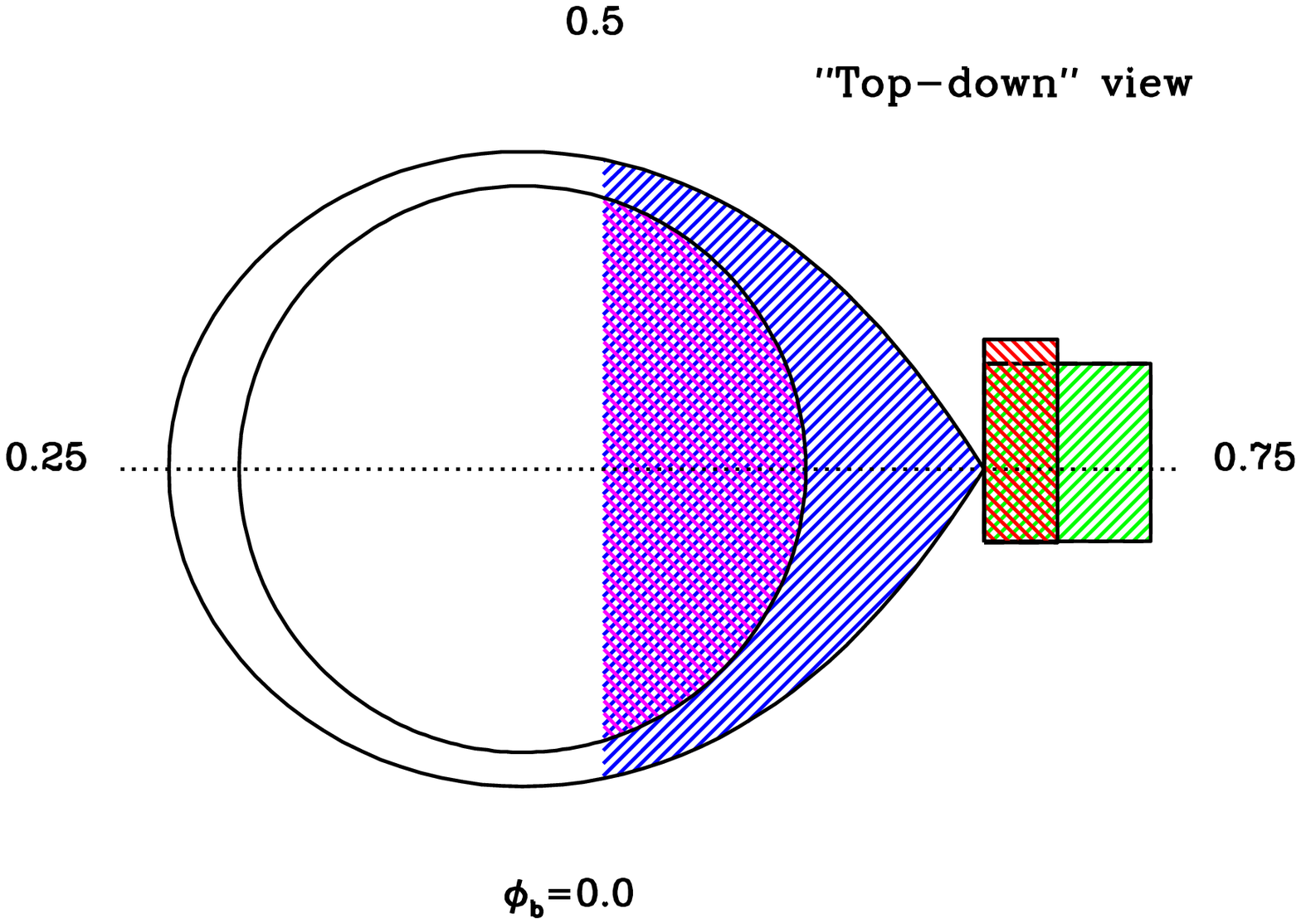}
\includegraphics[width=0.42\textwidth]{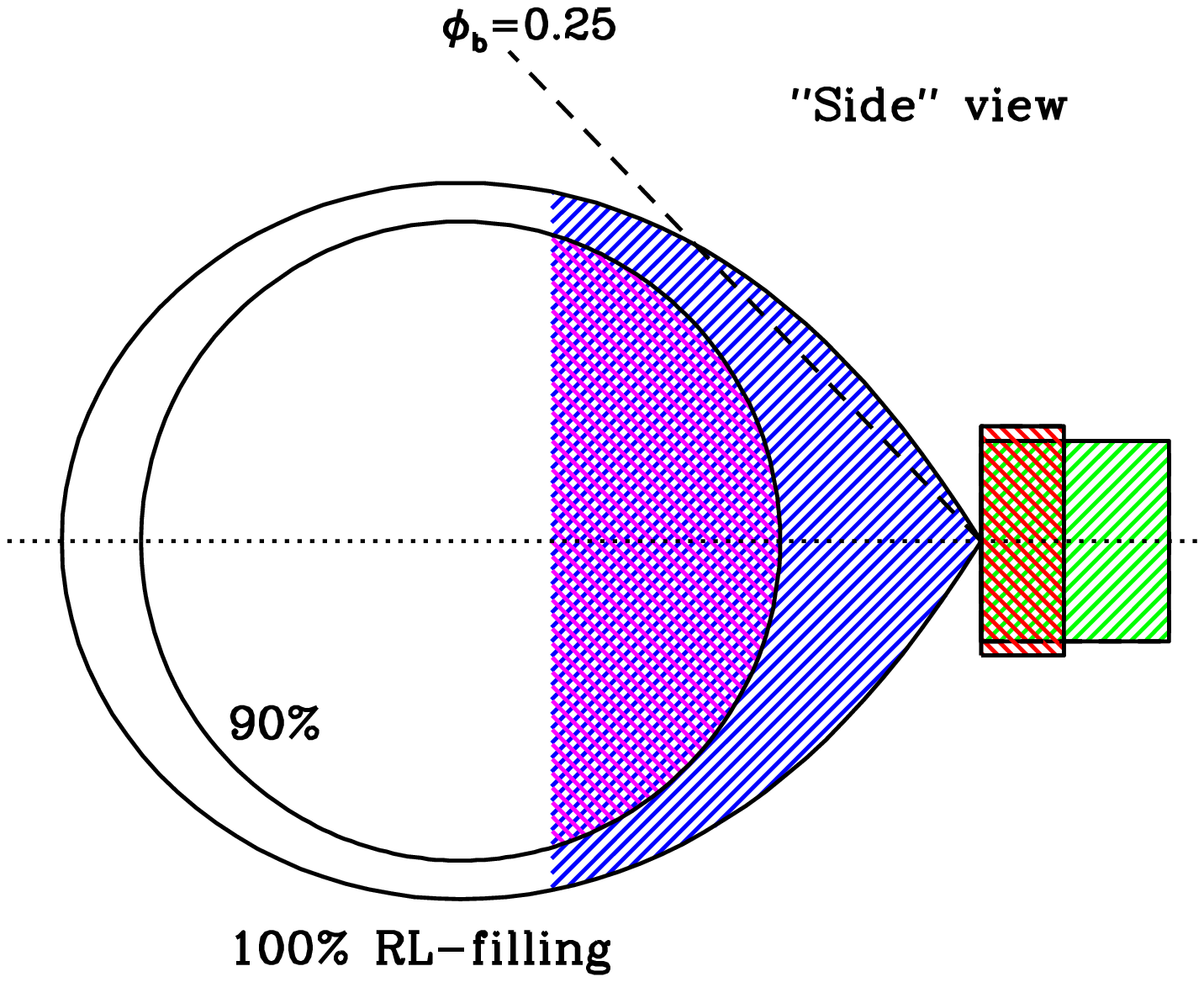}
\end{center}
\caption{(\textit{Top}) Schematic diagram illustrating the emission
  regions considered in the model. The orbital plane is in the plane
  of the page with the secondary star rotating counter-clockwise about
  the center of mass of the system. The labels mark the general
  directions to the observer at both conjunctions and quadratures.
  The blue area corresponds to X-ray emission from the face of a Roche
  lobe-filling secondary star. The magenta is for a companion that is
  90\% the size of its Roche lobe. The green and red areas are for
  isotropically emitting cylindrical regions at inner Lagrangian point
  (L1). Note the offset of the regions with respect to the semi-major
  axis in the direction opposite the sense of rotation of the star
  (\textit{dotted line}).  (\textit{Bottom}) Same as above but with
  the orbital plane perpendicular to the plane of the page. The dashed
  line shows the line of sight from the L1 point to the observer at
  superior conjunction ($\phi_b=0.25$) assuming $i=46^{\circ}$.}
\end{figure}

One possibility is that the pulsar wind is interacting with material
near the surface of the secondary star. If this is indeed the case,
the shape of the X-ray-emitting region would roughly correspond to the
shape of the side of the secondary star facing the pulsar (Figure
7). For this scenario, as illustrative examples, we consider both the
case of a Roche-lobe filling star and one that is 90\% the size of its
Roche lobe. It is apparent from Figure 8 that in both cases the
resulting lightcurves (blue and magenta lines, respectively)
approximate the shape of the observed modulations fairly well.


%
%
\begin{figure}[!t]
\begin{center}
\includegraphics[width=0.42\textwidth]{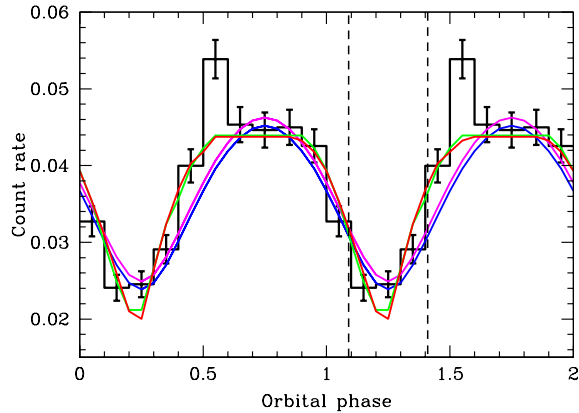}
\caption{Resulting model lightcurves fitted to the observed
  \textit{Chandra} lightcurve in the 0.3--8 keV band. The colors
  correspond to the shaded regions in Figure 7. The dashed lines mark
  the orbital phase range at which the L1 point is eclipsed by the
  secondary star for the Roche lobe filling case.  In all instances,
  the nominal orbital parameters of the J1023+0038 system are assumed.
  The bin with the maximum count rate at $\phi_b=0.5-0.6$ has been
  excluded from the lightcurves fits.}
\end{center}
\end{figure}

An alternative origin of the observed X-rays is interaction of the
pulsar wind with matter issuing from the Roche-lobe filling companion
through the L1 point \citep[see, e.g., Figure 2 in][]{Bog05}. As an
approximation to this outflow, we take a cylindrical volume with one
of the circular bases centered on the L1 point (Figure 7).  In this
configuration, it is possible to account for an eclipse minimum
occuring before $\phi_b=0.25$ (as suggested by the data, albeit at
marginal significance) by considering a region that is offset away
from the direction of motion of the secondary star in the binary. Two
representative examples of such a ``stream'' geometry that reproduce
the rough shape of the \textit{Chandra} lightcurve are shown in Figure
8 (green and red lines).

We note that it is also very possible that both emission from the
stellar surface and from a stream at L1 contribute to the observed
non-thermal X-ray flux. However, due to the limited photon statistics,
substantially deeper observations are required to unambiguously
determine the shock geometry by way of lightcurve
modeling. Nonetheless, Figure 8 illustrates that an intrabinary shock
localized primarily near L1 and/or at the face of the companion is a
very plausible interpretation of the observed X-ray flux modulations.

\section{Constraining the Physics of the Intrabinary Shock}

The purely geometric model employed above pays no regard to the actual
physical mechanism responsible for the production of the observed
X-ray emission. Nevertheless, the geometric information obtained can be
used to place constraints on the physics of the shock.  Given the
small separation between the pulsar and the companion
($1.2\times10^{11}$ cm), the shock must occur in a relatively strong
magnetic field. Therefore, synchrotron emission from accelerated
charged particles is the likely X-ray production mechanism.  The X-ray
luminosity in the shock is dependent on both the postshock magnetic
field strength and the wind magnetization factor $\sigma$, defined as
the ratio between the Poynting flux and particle flux (or,
equivalently, the ratio of magnetic and particle energy
densities). Since $\sigma$ at the shock front is not known \textit{a
  priori}, we must consider both a wind dominated by kinetic energy
($\sigma = 0.003$ as seen in the Crab nebula) and one that is
magnetically dominated ($\sigma \gg 1$). Following the formalism
described by \citet{Arons93}, the magnetic field strength immediately
upstream of the shock is given by
$B_1=[\sigma/(1+\sigma)]^{1/2}(\dot{E}/cf_{p}r^2)^{1/2}$. Here $f_p$
is a geometric factor that defines the fraction of the sky into which
the pulsar wind is emitted, and $r$ is the distance between the MSP
and the shock. We take $r\approx1\times10^{11}$ cm, the distance from
the MSP to L$_1$ assuming $M_{\rm MSP}=1.4$ M$_{\odot}$ and
$i=46^{\circ}$.  For an isotropic pulsar wind ($f_{p}=1$) and
$\dot{E}=5\times10^{34}$ ergs s$^{-1}$ we obtain $B_1\approx 0.75$ G
($\sigma=0.003$) and $B_1\approx 13$ G ($\sigma\gg1$).  The implied
magnetic field strength beyond the shock is then $B_2=3B_1\sim 2.1$ G
or $B_2\sim39$ G.  Production of the observed
$\varepsilon=0.3-8\sim1$ keV photons via synchrotron emission requires
relativistic electrons/positrons with Lorentz factors
$\gamma=2.4\times10^5(\varepsilon/B_2)^{1/2}\approx10^5$.  The
corresponding synchrotron radiative loss time is $t_{\rm
  synch}=5.1\times10^8(\gamma B_2^2)^{-1}\sim10-700$ s \citep{Ryb79}.
The geometric model presented earlier indicates that the shock region
is of order the diameter of the secondary star
($r\approx5\times10^{10}$ cm). From here, the residence times of the
radiating particles in the shock region are $t_{\rm
  flow}=c/3r\approx5.1$ s (for $\sigma=0.003$) and $t_{\rm
  flow}=c/r\approx1.7$ s (for $\sigma\gg 1$).

The expected synchrotron luminosity from the shock can be estimated
from $f_{\rm shock} \Delta\varepsilon L_{\varepsilon}=f_{\rm
  synch}f_{\gamma}f_{\rm geom}\dot{E}$, where $f_{\rm synch}$ is the
radiative efficiency of the synchrotron emission, $f_{\gamma}$ is the
fraction of the intercepted spin-down energy flux that goes into
accelerating electrons with Lorentz factor $\gamma$, $f_{\rm geom}$ is
the fraction of the pulsar wind that interacts with matter from the
companion, $f_{\rm shock}$ is the portion of the total observed X-ray
luminosity that originates from the intrabinary shock, and
$L_{\varepsilon}$ is the synchrotron luminosity over the energy band
$\Delta\varepsilon$. For simplicity, we will assume
$\Delta\varepsilon\approx1$ keV centered at 1 keV.  Since $f_{\rm
  synch}=(1+t_{\rm synch}/t_{\rm flow})^{-1}$, we obtain $f_{\rm
  synch}\approx0.007$ and $f_{\rm synch}\approx0.16$ for
$\sigma=0.003$ and $\sigma\gg 1$, respectively.  The companion star
intercepts $f_{\rm geom} \approx 0.01$ of the pulsar's outflow for the
case of an isotropic wind. It is important to note, however, that the
MSP outflow most likely exhibits significant anisotropy, with the bulk
of the wind emitted equatorially \citep[as seen in the Crab
  pulsar;][]{Hest95,Mich94}. Since the spin axis of the pulsar has
likely aligned with the orbital angular momentum vector during the
LMXB accretion phase \citep{Bhatt91}, the wind of PSR J1023+0038 should thus be
preferentially emitted in the orbital plane.
In the extreme case of a sheet-like wind confined entirely to the
orbital plane, the secondary star would receive $f_{\rm
  geom}\approx0.07$ of the pulsar wind's power. Based on this, we set
$0.01\lesssim f_{\rm geom}\lesssim 0.07$.

Using the count rate maxima predicted by the model lightcurves shown
in Figure 8, the intrinsic synchrotron luminosity of the shock in the
absence of eclipses is estimated to be
$L_{\varepsilon}\approx3\times10^{31}$ ergs s$^{-1}$ for particles
with $\gamma\approx10^5$ in a band $\Delta\varepsilon=1$ kev centered
on $1$ keV. With $f_{\rm shock}\approx0.87$ as deduced from the
spectral fits, we obtain $0.93\lesssim f_{\gamma}\lesssim5.6$
($\sigma=0.003$) and $0.04\lesssim f_{\gamma}\lesssim0.25$
($\sigma\gg1$).  For $\sigma=0.003$, the derived $f_{\gamma}$ for an
isotropic wind exceeds unity and even in the highly anisotropic case
is unrealistically high. The same applies if we consider a factor
of $2-3$ larger emission region and stronger upstream
magnetic field.  Since it is likely that $f_{\gamma}\ll1$ (e.g.,
$f_{\gamma}=0.04$ for the Crab pulsar and nebula), the unphysical
values of $f_{\gamma}$ for $\sigma=0.003$ suggest that at the distance
of the shock, the wind of PSR J1023+0038 is magnetically dominated
($\sigma\gg1$). Furthermore, for $\sigma\gg1$, a value of $f_{\gamma}$
comparable to that of the Crab pulsar requires significant anisotropy
in the wind.

It is interesting to highlight the apparent enhancement of emission at
eclipse egress ($\phi_b=0.5-0.6$) since it likely arises due to
Doppler beaming in the direction of the post-shock flow
\citep{Arons93}. Similar, albeit much less statistically significant
features have been seen in other eclipsing MSP systems \citep[see,
  e.g.,][]{Stap03,Bog05}. For PSR J1023+0038, the implied flow
direction is opposite to the sense of motion of the secondary star in
its orbit, indicating that the combined effects of the gravitational,
pulsar wind pressure, and Coriolis forces together with the outflow
velocity relative to the orbital velocity, conspire to force the
outflowing gas towards the trailing edge of the eclipse \citep[see,
  e.g.,][]{Tav91}. This effect can also account for the absence of a
matching X-ray flux increase at ingress as it results in a weaker
outflow towards the leading edge of the eclipse. Within the available
photon statistics, the factor of $1.2-1.3$ increase in brightness at
$\phi_b=0.5-0.6$ relative to the flux outside of eclipse
($\phi_b\approx0.6-0.9$) is in agreement with the range $1.3-2.2$ predicted by
\citet{Arons93} based on the expected range of post-shock wind flow
velocities ($c/3-(c/3)^{1/2}$).

\section{CONCLUSION}

We have presented \textit{Chandra} ACIS-S observations of the PSR
J1023+0038 binary system. The data definitively confirm the X-ray flux
variability at the binary orbital period. A simple geometrical model
can account for the general shape of the X-ray lightcurve of the
binary, indicating that the observed variability is primarily due to a
partial geometric eclipse of the intrabinary shock by the secondary
star. The model suggests that the shock is likely localized near L1
and/or is formed by the interaction of the pulsar wind and matter at
the surface of the secondary star.  The energetics of the shock
emission favor a magnetically dominated ($\sigma\gg1$) pulsar wind at
the shock front. This is not surprising, given that the shock occurs
in a $\sim$$10^4$ times stronger magnetic field than in the case of
the Crab pulsar \citep[where the wind is dominated by kinetic energy;
  see][]{Ken84}. The X-ray data also favor a significantly
anisotropic pulsar wind that is preferentially emitted in the orbital
plane, which, in turn, requires close alignment of the pulsar spin and
orbital angular momentum axes, as expected from binary evolution.

Combined with the study presented in \citet{Arch10}, the spectral
continuum suggests a dominant non-thermal component and the presence
of a fainter thermal component that, at least in part, originates from
the hot polar caps of the pulsar. In this sense, in terms of its X-ray
properties, the pulsar itself appears to be similar to the majority of
``recycled'' MSPs detected in X-rays to date
\citep{Zavlin06,Bog06,Bog09,Bog11}. A portion of the thermal X-ray
emission may also come from an optically-thin thermal plasma, possibly
the same plasma responsible for the radio eclipses.

We find no evidence for a wind nebula associated with
the pulsar. The shape, luminosity, and spectrum of a PWN depend on the
angular distribution, magnetization and energy spectrum of the
unshocked wind, as well as on the pulsar velocity and the properties
of the ambient medium.  Therefore, the absence of a nebula down to a
fairly deep limit of $\lesssim$$3.6\times10^{29}$ ergs s$^{-1}$,
corresponding to $\lesssim$$7\times10^{-6}$ of the pulsar's $\dot{E}$,
implies that the combination of low ambient density, moderate space
velocity, and/or unfavorable wind geometry are insufficient to
drive an X-ray-luminous bow shock.




Owing to its proximity, PSR J1023+0038 is one of the best-suited
targets for studies of these rare ``transition'' systems and thus
warrants further X-ray observations, especially in combination with
optical spectroscopy and $\gamma$-ray variability study
\citep[see][for the recent detection of this system with the
  \textit{Fermi} Large Area Telescope]{Tam10}.  Constraining the
properties of the intrabinary shock in J1023+0038 and similar systems
has important implications for understanding the X-ray properties of
accreting MSPs in quiescence, such as the prototypical SAX
J1808.4--3658 \citep{Wij98}. It has been suggested that the
non-thermal X-ray radiation observed from certain quiescent low-mass
X-ray binaries is also due to an active rotation-powered wind but this
cannot be directly confirmed due to the absence of radio pulsations
\citep{Camp04}. Alternative scenarios such as on-going, low-level
accretion \citep[see, e.g.,][and references therein]{Gar01} or a
propeller regime \citep[][and references therein]{Wij03} have also
been proposed as explanations for the non-thermal emission.  An X-ray
spectroscopic and variability study analogous to that presented above
could reveal the signature of an interacting rotation-powered pulsar
wind in such systems, thereby providing further evidence for the
evolutionary relation of LMXBs and MSPs.

\acknowledgements We thank C.-Y.~Ng for helpful discussions. The work
presented was funded in part by NASA \textit{Chandra} grant GO0-11065X
awarded through West Virginia University and issued by the
\textit{Chandra} X-ray Observatory Center, which is operated by the
Smithsonian Astrophysical Observatory for and on behalf of NASA under
contract NAS8-03060. S.B. is supported in part by a CIFAR Junior
Fellowship. A.M.A. is supported by a Schulich graduate
fellowship. V.M.K. acknowledges support from NSERC, FQRNT, CIFAR, a
Killam Research Fellowship, and holds a Canada Research Chair and the
Lorne Trottier Chair in Astrophysics and Cosmology. J.W.T.H. is a Veni
Fellow of the Netherlands Foundation for Scientific Research
(NWO). I.H.S. is supported in part by a NSERC Discovery Grant.  This
research has made use of the NASA Astrophysics Data System (ADS) and
software provided by the Chandra X-ray Center (CXC) in the application
package CIAO.

Facilities: \textit{CXO}

\end{document}